\documentstyle[prb,aps,preprint,epsf]{revtex}
\tighten
\begin{document}
\tightenlines
\title{Anomalous Thermal Transport in Quantum Wires}
\author{Rosario Fazio$^{(1)}$, F.W.J. Hekking$^{(2,3)}$, and D.E. 
        Khmelnitskii$^{(2,4)}$}
\address{$^{(1)}$Istituto di Fisica, 
Universit\'a di Catania, \& INFM, Viale A. Doria 6, I-95129 Catania, Italy\\
$^{(2)}$Cavendish Laboratory, University of Cambridge, Madingley Road, 
Cambridge CB3 0HE, United Kingdom\\
$^{(3)}$Theoretische Physik III, Ruhr-Universit\"at Bochum, 44780 Bochum,
Germany\\
$^{(4)}$L.D. Landau Institute for Theoretical Physics, 117940 Moscow, 
Russia\\}
\date{\today}
\maketitle

\begin{abstract}
We study thermal transport in a one-dimensional quantum wire, connected 
to reservoirs.  Despite of the absence of electron backscattering, interactions
in the  wire strongly influence thermal transport. 
Electrons  propagate with unitary transmission
through  the wire and electric conductance is not affected. 
Energy, however, is carried by bosonic excitations (plasmons)
which  suffer from scattering even on scales much larger than the
Fermi wavelength. If the electron density varies
randomly, plasmons are localized and  {\em charge-energy separation} 
occurs.  We
also discuss the effect of plasmon-plasmon interaction using Levinson's theory
of nonlocal heat transport.
\end{abstract}


\narrowtext
In the Fermi liquid theory of thermoelectric transport,  both charge and energy
are carried by fermionic  quasiparticles~\cite{Abrikosov88}.  This manifests
itself in a universal relation  between the electric conductivity $\sigma$ and
thermal conductivity $\kappa$,  known as the Wiedemann-Franz (WF) law ($k_B =
\hbar =1$),
\begin{equation}
\frac{\kappa}{\sigma T} =\frac{ \pi^2}{3e^2} \;\; .
\label{WF}
\end{equation}
The validity of Eq.~(\ref{WF}) has been confirmed in the case of arbitrary 
impurity scattering~\cite{Chester61} and in the presence of electron-electron 
interactions~\cite{Castellani87} within the Fermi liquid approach.

At low temperatures, in conductors of small size, phase-coherent electron
propagation dominates transport. Mesoscopic contributions to thermoelectric 
coefficients in the diffusive regime are quite 
significant~\cite{AnisovichLesovik}. 
The thermoelectric coefficients of a ballistic quantum point contact have  been
studied experimentally~\cite{MolenkampDzurak} and 
theoretically using a scattering  approach~\cite{Sivan86,Butcher90,VanHouten92}.
In this case, the observed violations of the WF law can be attributed to the
strong energy dependence of the scattering matrix.

Electron-electron interactions in low-dimensional systems may lead to non  Fermi
liquid behavior. In this context transport properties of quantum  wires are of
considerable current interest~\cite{Tarucha95Yacoby96}.   Theoretically these
systems are studied in the framework of the Luttinger  liquid  (LL)
model~\cite{Haldane}. Revived interest in the transport in LLs was triggered by
the work of Kane and Fisher~\cite{Kane92}. The effect of  interaction on the
electric conductance crucially depends on the way in  which the quantum wire is
connected to the measuring leads  (reservoirs)~\cite{Maslov95Ponomarenko95Safi95}.
More
recently thermal transport in a  LL was considered~\cite{Kane96} and deviations
from the Fermi liquid  relation, Eq.~(\ref{WF}), were predicted.

\begin{figure}
{\epsfxsize=14cm\epsfysize=4cm\epsfbox{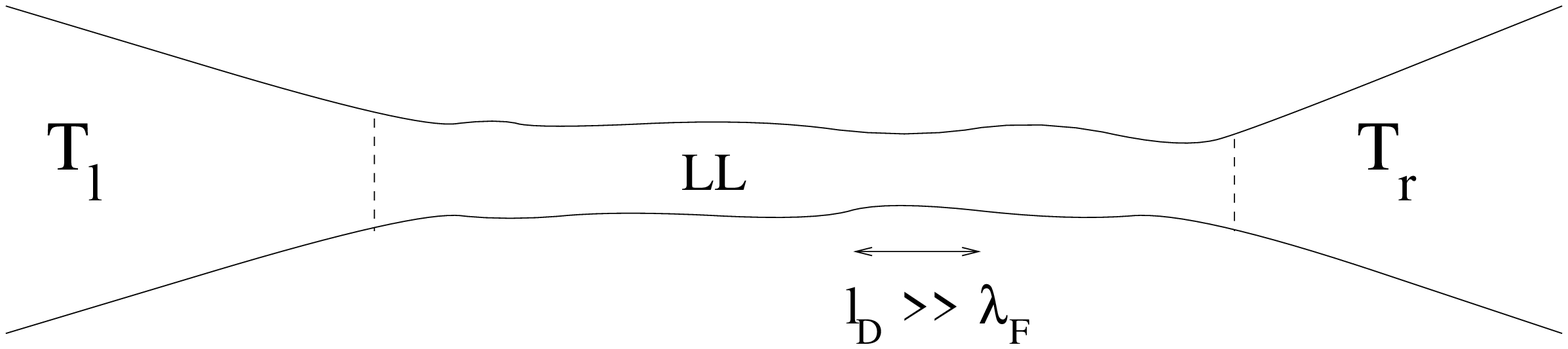}}
\caption{The 1D wire, connected adiabatically to two reservoirs
kept at different temperatures. The wire has some inhomogeneities on a 
scale which is much larger than $\lambda_F$.}
\label{fig1}
\end{figure}

The low energy excitations of an interacting one-dimensional (1D) system are
long wavelength density oscillations (plasmons) which have bosonic  character.
This has drastic consequences for thermoelectric transport, since the scattering
properties of the electrons (which determine the  transfer of charge) are in
general quite different from the scattering  properties of the plasmons
(responsible for the transfer of energy).  In this paper we study this distinct
difference by considering situations in which the dc electric conductance is not
affected by interaction  in contrast to the thermal conductance. Specifically we
consider a one channel LL with spatially varying density, which is connected to
two  reservoirs as it is shown in Fig.~\ref{fig1}.  If the spatial variation
related to these inhomogeneities occurs  on a length scale much larger than the
Fermi wavelength $\lambda_F$, electrons will {\em not  suffer any
backscattering}.  The electric DC conductance will therefore be  given by the
universal value $G= e^2/2\pi$. However, plasmons with wavelengths 
much larger than $\lambda _{F}$ will suffer backscattering.
Under these circumstances the thermal conductance is strongly affected 
by interactions.

The Hamiltonian which describes such an inhomogeneous electron liquid can be 
written as~\cite{Gramada97} $H=\int dx {\cal H}(x)$, where
\begin{equation}
{\cal H}(x)
=
\frac{p^{2}}{2mn(x)} 
+ \frac{1}{2} \left[ V_{0} + \frac{\pi ^{2}}{m}n(x)\right]
\left( \partial _x n u\right) ^{2} \; .
\label{ham}
\end{equation}
Here, $u(x)$ and $p(x)$ are the displacement of the electron liquid and the 
conjugate momentum, respectively, satisfying $[u(x),p(x')] = i \delta (x-x')$.
The local average electron concentration is $n(x)$. Furthermore, $V_{0}$ is the
interaction strength (the zero-momentum  Fourier component of the interaction
potential). The connection of a 
quantum wire of length $d$ to reservoirs 
is modeled by
a space-dependent interaction 
$V_{0}(x)$ ($V_{0}(x) = V_{0}$ if $0 < x < d$ and zero 
otherwise)~\cite{Maslov95Ponomarenko95Safi95,footnote}.
For later use we define the parameter
$g = (1+V_{0}/\pi v_{F})^{-1/2}$, characterizing the interaction 
strength in the wire. 
We furthermore take $n(x) = n_{0} \equiv m v_F /\pi$ in the reservoirs 
({\em i.e.} for $x< 0$ or  $ x>d $.
Using the continuity equation 
$\partial _x J_E(x) + \partial _t {\cal H}(x)=0$,
the energy current $J_E(x)$ can be expressed
in terms of $u(x)$ and $p(x)$, 
\begin{equation}
J_E(x)
=
-
\frac{1}{2m} \left[V_{0} + \frac{\pi ^{2}}{m}n(x)\right]
\left\{ p(x),\partial _x (nu)\right\} \;\; ,
\label{encur}
\end{equation}
where $\{ \ldots, \ldots \}$ denotes the anticommutator. We will show that the
calculation of the average energy current amounts  to the solution of a well
defined scattering problem in analogy to the well known Landauer-B\"uttiker
approach for quantum  transport in noninteracting electron
systems~\cite{Buettiker92}.

To this end, we diagonalize the  Hamiltonian~(\ref{ham}) using a basis of 
scattering  states $\phi _{k,\alpha}(x)$ which satisfy the wave equation
\begin{equation} 
- \hat{h} \phi _{k,\alpha}
 =
\omega _{k}^{2} \phi _{k,\alpha}(x) \;\; ,
\label{waves}
\end{equation}
where $\omega _{k}$ is the energy of the given mode and 
$$
\hat{h} = 
\sqrt{n(x)} \partial_x \left( V_{0}/m + \pi^{2}n(x)/m^2 \right)
\partial_x \sqrt{n(x)}\;\; .
$$
The index $\alpha=r,l$ labels the  states incident  
from the right and left
reservoir, respectively, with wavevector $k$. The scattering 
states have asymptotics
\begin{equation}
\begin{array}{ll}
\phi _{k,l}(x)
&=
e^{i k x} + r_{\omega _{k}} e^{-ik x} \mbox{ for } 
x \to -\infty ,\\
&= t _{\omega_{k} }e^{ikx} \mbox{ for } x \to \infty ;\\
\phi _{k,r}(x)
&=
t _{\omega_{k} }e^{-ik x} \mbox{ for } x \to -\infty ,\\
&=
e^{-ik x} - r_{\omega_{k}}^{*}(t_{\omega_{k}}/t_{\omega_{k}}^{*}) 
e^{ik x} \mbox{ for } x \to \infty ,
\end{array}
\label{asymp}
\end{equation}
where $r_{\omega _{k}}$ and $t_{\omega_{k}}$ are the reflection and 
transmission amplitude, respectively. 
We emphasize that
$\phi _{k,\alpha}$ describe {\em  plasmon waves} rather than
electronic excitations.  The diagonalized Hamiltonian is given by
\begin{equation}
H
= 
(1/2) \sum _{\alpha} \int dk
\omega _{k} (b^{\dagger} _{k,\alpha} b_{k,\alpha} + 
b_{k,\alpha} b^{\dagger} _{k,\alpha}) .
\label{nboseham}
\end{equation}
Here, the operators $b$ and $b^{\dagger}$ obey the Bose commutation
relation  
$
[b_{k,\alpha}, b_{k',\alpha'} ^{\dagger} ]= \delta _{\alpha, \alpha'} 
\delta (k - k') 
$.
The displacement field and its conjugate momentum can be expressed as
$$
p(x) 
= 
\sum \limits_{\alpha} \int \limits_0^{\infty} dk 
\frac{1}{i} \sqrt {\frac{\Omega_{k}(x)}{2}}
\left[ \phi _{k,\alpha}(x) b_{k,\alpha} - \phi _{k,\alpha}^{*}(x) 
b^{\dagger} _{k,\alpha}\right] ,
$$
\begin{equation}
u(x) 
 = 
\sum \limits_{\alpha} \int \limits_0^{\infty}dk 
\sqrt {\frac{1 }{2 \Omega_{k}(x) }} 
\left[ \phi _{k,\alpha}(x) b_{k,\alpha} + 
\phi _{k,\alpha}^{*}(x) b^{\dagger} _{k,\alpha}\right] ,
\label{nbosescat}
\end{equation}
where $\Omega_{k}(x) = m n(x)\omega _{k}$.
At this point we want to stress that the above formulation is applicable 
because each mode contributes independently to the energy flux. A similar
approach cannot be applied to determine the electric current, as the charge is
transported through the sample via a complicated  superposition of plasmon modes.

Now we are in a position to calculate the average energy current.
This quantity is space-independent because of current conservation,
hence we calculate it, say, for $x \to \infty$. We substitute the appropriate 
asymptotics from~(\ref{asymp}) into~(\ref{nbosescat}), then perform 
the Gibbs average of the anticommutator in~(\ref{encur}) with respect 
to the Hamiltonian~(\ref{nboseham}). 
As a result we find
\begin{equation}
\langle J_E \rangle = \frac{1}{2\pi} \int \limits_0^{\infty} d\omega
\mid t_{\omega} \mid^2 \left[ n_l(\omega) - n_r(\omega)\right] \;\; ,
\label{J_E}
\end{equation}
where $n_{\alpha}(\omega)$ is the Bose function of 
reservoir $\alpha$. This result holds as long as electrons are not
backscattered by inhomogeneities.
In the linear response regime, 
when the temperature difference $\Delta T$ between the reservoirs is vanishingly 
small,
the thermal conductance is readily evaluated to be
\begin{equation}
K
=
\frac{1}{8\pi T^{2}} \int \limits _0^ \infty d\omega 
\frac{\omega ^{2}}{\sinh ^{2}(\beta \omega/2)}
\mid t_{\omega} \mid ^2 ,
\label{K}
\end{equation}
where $\beta$ is the inverse temperature $1/T$.
For noninteracting electrons, $\mid t_{\omega} \mid = 1$, this expression reduces
to the well-known result $K= \pi T /6$. Eq.~(\ref{K}) shows that inhomogeneities 
strongly affect
thermal transport in an interacting quantum wire, even in the absence of any
electron backscattering. This is essentially the analogue of the Kapitza
boundary resistance~\cite{Swartz89}. Below we study the behavior of 
$K$, Eq.~(\ref{K}), in two relevant limits.

\begin{figure}
{\epsfxsize=14cm\epsfysize=10cm\epsfbox{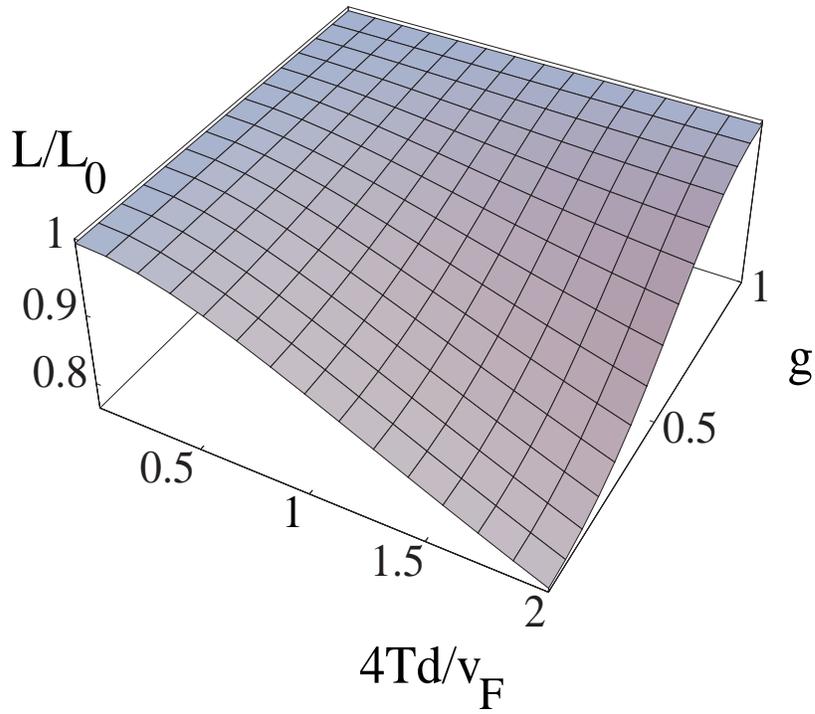}}
\caption{The Lorentz number for an ideal wire attached to reservoirs is 
plotted as a function of the temperature and the interaction strength.}
    \label{fig2}
\end{figure}

(i) The simplest situation occurs when an interacting wire of finite length  with a
constant electron density $n_0$ is connected adiabatically
to two noninteracting reservoirs. In this case the solutions of
Eq.~(\ref{waves}) inside the wire are plane waves, with 
momentum $g k$. 
The plasmon transmission coefficient can be calculated
explicitly; it is strongly frequency-dependent with characteristic
frequency $v_F/gd$
due to the mismatch of momenta at $x=0$ and $x=d$. 
As a result, $K$ will be suppressed
below its noninteracting value. The Lorentz number
$ L=K/GT = 2\pi K/e^{2} T$ as a function of temperature and  interaction strength is
plotted in Fig.~\ref{fig2}. In the low-temperature limit, the Lorentz number
attains its noninteracting value $ L_0 = \pi^2/(3e^2)$ (since the zero
frequency transmission coefficient equals unity) and decreases with increasing
temperature and interaction strength. Notwithstanding the  fact that a
decrease of $ L$ is a genuine feature of the system, the  actual
quantitative suppression depends on the specific choice of the space  dependence
of the interaction strength. Note, finally, the profound difference 
between our results and those obtained in~\cite{Kane96}. 
For an infinitely long interacting wire without reservoirs, 
the electric conductance is renormalized~\cite{Kane92}. However, the thermal 
conductance is unaffected by interactions and an
enhanced Lorentz number signals the  breakdown of Fermi liquid 
theory~\cite{Kane96}. In the presence of reservoirs, the 
electric conductance is unrenormalized~\cite{Maslov95Ponomarenko95Safi95} but the thermal 
conductance is suppressed, leading to a suppression of $ L$.

\begin{figure}
{\epsfxsize=14cm\epsfysize=10cm\epsfbox{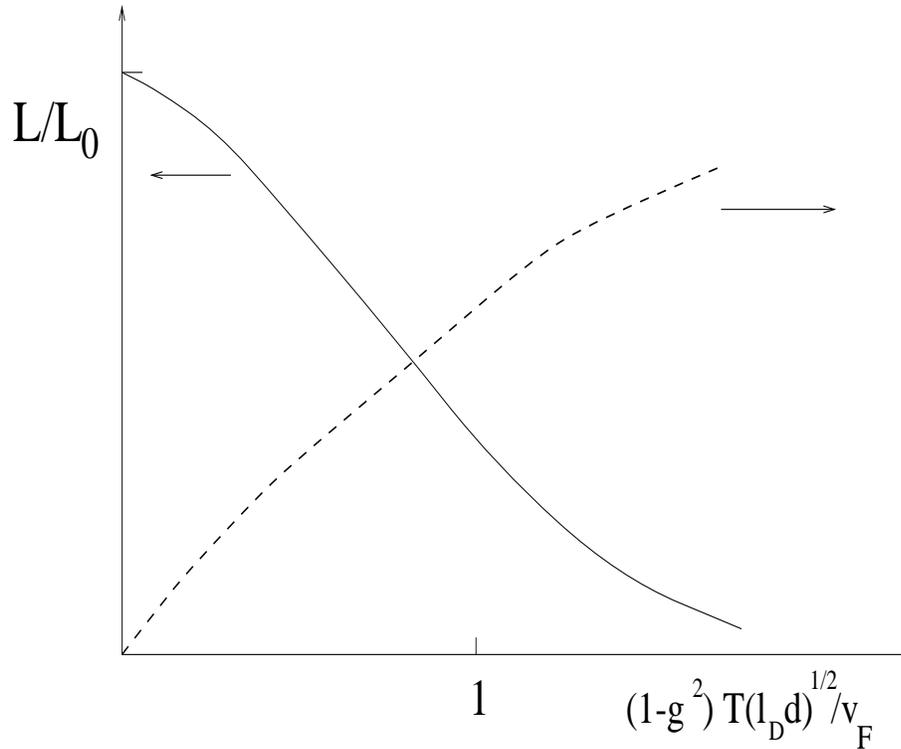}}
  \caption{The behavior of the Lorentz number and the thermal conductance is
sketched for the case in which the wire is disordered. The saturation of 
$K$ at some value $K_{0}$ is related to the localization of the high frequency 
plasmons. }
    \label{fig3}
\end{figure}

(ii) Even in the best samples, random variations of electron density on scales
$l_D$ much larger than $\lambda_F$ are unavoidable. Plasmons, therefore, 
propagate in a random medium and, depending on their energy, can be  
localized. We anticipate that in some temperature range charge and energy 
are spatially separated ({\em charge-energy separation}).
In order to model the randomness, we decompose the electron density as
$n(x) = n_0 +
\delta n(x)$  where the random component has a normal distribution with variance
$$
\langle \delta n(x) \delta n(y) \rangle 
=
n_0^2 \delta \left(\frac{x-y}{l_D}\right).
$$
We calculated the average plasmon transmission coefficient with the help of the 
invariant embedding technique developed in Ref.~\cite{Doucot87}. The decay of
the plasmon wave inside the interacting wire is governed by the lengthscale
\begin{equation}
\xi_{\omega} = \frac{v_F}{g} \tau_D(\omega) =
                \frac{2\pi^2 }{g^6}
                \frac{v_F^2}{V_0^2}
                \frac{v_F^2}{l_D\omega^2},
\label{xi}
\end{equation} 
where we introduced the lifetime for impurity scattering
$\tau_D(\omega)$. Eq.~(\ref{xi}) was derived using the Golden Rule in
Ref.~\cite{Gramada97}. The transmission coefficient asymptotically decays as
$$
\mid t_{\omega} \mid ^2 \sim e^{-d/\xi_{\omega}}
$$
for $d \gg \xi_{\omega}$. At low enough temperatures, 
$T < \omega ^*_1 = (2\pi^2 v_F^4/g^6l_DV_0^2d)^{1/2}$, the thermal plasmons
propagate ballistically. In the opposite limit $T > \omega ^*_1$, localization
of high-frequency plasmons occurs and the thermal  conductance rapidly saturates
to some constant value $K_{0}$ as sketched in Fig.~\ref{fig3}.

A finite plasmon lifetime is not only caused by scattering off inhomogeneities.
The nonzero curvature of the single electron spectrum leads to interactions
between plasmons. 
The effect of these two scattering mechanisms on the plasmon kinetics 
is very different, as we will discuss in the remainder of this paper.
Scattering off inhomogeneities relaxes the momentum
but  does not lead to thermal equilibrium. 
 This is established by  plasmon
interactions which, in lowest order, are described by the cubic
Hamiltonian~\cite{Haldane}
\begin{equation}
H_{\rm int}
=
-\frac{1}{2}\int dx \frac{p^{2}}{mn(x)}\partial_x(nu)
-\frac{\pi^{2}}{6m} \int dx \left( \partial_xn_{0}u\right)^{3} \;\; .
\label{cubicham}
\end{equation}
The rate for three-plasmon scattering can be calculated using the Golden Rule. 
Special care is needed, however, in this case. Since the dispersion  relation is
linear in $k$, momentum and energy conservation are simultaneously satisfied,
hence the rate is infinite. In the presence of impurity scattering it can be
regularized, because a state with given energy $\omega$
corresponds to a wave packet with $\langle k \rangle = g\omega /v_{F}$
and a width $\langle (\delta k)^{2} \rangle \sim (g/v_F\tau_D(\omega))^2$. 
Using the broadened dispersion $\omega _k = v_F\mid k\mid /g  + 
i\tau_D^{-1}(v_Fk/g)$, we find the rate associated with
spontaneous decay of a plasmon
\begin{equation}
\frac{1}{\tau (\omega)}
=
\frac{(3+g^2)}{16g^2}\frac{\omega ^2}{n_0^2 V_0^2}\frac{v_F}{l_D}\;\; .
\label{srate}
\end{equation}
Interactions are important for plasmons with energies larger than the cross-over
frequency $\omega ^*_2 \sim v_F/\sqrt{d \lambda _F}$, since they experience
interactions while diffusing over distances of the order of the length of the
wire $d$. The ratio $(\omega ^*_2/\omega ^*_1 )^2 \sim (V_0/v_F)^2
l_D/\lambda _F$ should be much larger than unity.
This, together with the condition 
$\langle (\delta k)^{2} \rangle \ll \langle k \rangle^2$ for the wave 
packet discussed above, implies
that $(\lambda_F/l_D)^{1/2} \ll V_0/v_F \ll 1$.

The scattering approach of thermal transport,
resulting in Eq.~(\ref{J_E}), applies when both reservoirs are kept at a
temperature smaller than $\omega^*_2$. At temperatures larger than $\omega^*_2$
local equilibrium tends to be established.
The difference between these two regimes can be distinguished, {\em 
e.g.}, by
considering the nonlinear response to a large temperature difference $\Delta T$
between the reservoirs. Using the scattering approach Eq.~(\ref{J_E}) one finds
\begin{equation}
J_E \sim
\Delta T \mbox{ if } \Delta T < \omega^*_2 .
\label{J_Esmall}
\end{equation}
If $\Delta T \gg \omega^*_2$, interaction between
plasmons can no longer be ignored. Using Levinson's
theory~\cite{Levinson80,Levinson86} of phonon heat transport in disordered
dielectrics and semiconductors, we will show that far from equilibrium the
energy current depends algebraically on the temperature difference $\Delta T$.

Below we closely follow the derivation given in
Ref.~\cite{Levinson80}. Since low energy plasmons propagate easier through a
disordered wire than high energy ones, the plasmon distribution function is
depleted at energies smaller than a certain cross-over frequency $\bar{\omega}$.
Therefore, in the presence of plasmon interactions, down energy conversion
occurs from thermal to low energy plasmons. The latter then carry the energy to
the cold end of the wire. The relevant time scale for down conversion 
is  
\begin{equation}
\frac{1}{\tau_B (\omega)}
=
\frac{2}{\tau (T)} \frac{\omega}{T}
=
 \frac{(3+g^2)}{8g^2}\frac{\omega T}{n_0^2 V_0^2}\frac{v_F}{l_D} .
\end{equation}
Comparing $\tau _{B}$ with the diffusion time $gd^{2}/v_{F}\xi_{\omega}$ 
one finds
\begin{equation} 
\bar{\omega} = T \left(\frac{d}{L_T}\right)^{-2/3}
\end{equation}
where $L_T= \{4\pi n_0 v_F^2/[g^5(3+g^2)]^{1/2}\}T^{-2}$. 
For frequencies below 
$\bar{\omega}$ the depleted distribution function obeys the stationary 
diffusion equation 
$(v_{F}\xi _{\omega}/g) \partial _{x}^{2} n = 1/\tau(T)$, hence it
is given 
by
\begin{equation} 
n(\omega) \sim \left(\frac{d}{L_T}\right)^{2}\left(\frac{\omega}{T}\right)^{2}.
\end{equation}
For frequencies larger than $\bar{\omega}$
the thermal distribution $n(\omega) \sim T/\omega$ is recovered.
Using this distribution function it is possible to evaluate the nonlinear 
energy current $J_E \sim \int d\omega \omega \xi _\omega n(\omega)/d$,
\begin{equation}
J_E \sim (\Delta T)^{4/3} \mbox{ if } \Delta T > \omega^*_2 .
\label{J_Elarge}
\end{equation}
The power law given in Eq.~(\ref{J_Elarge}) is {\em universal}: it does not
depend on the interaction strength. Still, this algebraic behavior is a genuine effect
of the electron-electron interaction: in the noninteracting case,
plasmons are ballistic at all frequencies and Eq.~(\ref{J_Esmall}) 
applies. The non-linearity in Eq.~(\ref{J_Elarge}) is a result of the 
interplay between electron-electron interactions and plasmon-plasmon 
interactions, which provides a mechanism for redistribution of energy 
over the plasmon spectrum.

{\bf Acknowledgments} We thank Ya.M. Blanter, K. 
Samokhin and especially E. Paladino for valuable discussions. 
We acknowledge the financial support of INFM (PRA-QTMD) and the 
European Community (Contract ERB-CHBI-CT941764).

\newpage

\end{document}